# How droplets dry on stretched soft substrates

Yixuan Du[a], Elmar Bonaccurso[b], Jianwei Guo[c], Kai Uhlig[a], Longquan Chen[d],

Binyu Zhao[a,*] & Günter K. Auernhammer[a]

[a] Leibniz Institute of Polymer Research Dresden, Dresden 01069, Germany
[b] Airbus Central R&T, Materials, Munich 81663, Germany
[c] School of Mechanics and Aerospace Engineering, Southwest Jiaotong University, Chengdu 610031, China
[d] School of Physics, University of Electronic Science and Technology of China, Chengdu 610054, China

* To whom correspondence should be addressed. E-mail: binyuzhao2014@sina.com.

**Abstract**

Liquid droplets usually wet smooth and homogeneous substrates isotropically. Recent research works have revealed that droplets sit, slide and spread anisotropically on uniaxially stretched soft substrates, showing an enhanced wettability and contact line mobility along the stretching direction. This phenomenon arises from the anisotropic deformation of the substrate below the contact line. Here, we investigate how the stretching of soft substrates affects droplets drying. We observe that water droplet evaporates with an elongated non-circular contact line on the stretched substrates and switches the elongation direction during evaporation. The contact line velocity and its temporal evolution depend on the orientation of the contact line relative to the stretching direction. On the substrate stretched by 250%, the contact line recedes about 20% of the droplet lifetime earlier along the stretching direction and faster than its perpendicular direction. When nanoparticles are added into the liquid, the circular deposition pattern, i.e., the so-called coffee-ring, becomes elongated along the direction perpendicular to the stretching direction. Particularly, such non-circular deposition pattern exhibits periodic height gradients along its rim. The finer structure of the pattern can be controlled by applying different stretching ratios to the soft substrate and thus are correlated to the anisotropic surface stresses near the contact line.

**Significance:** Droplets evaporation on solid substrates is a ubiquitous phenomenon and relevant in many natural and industrial processes. Well known is the coffee-ring phenomenon. Many studies have succeeded in promoting, suppressing or even reversing the formation of the coffee-ring by using non-spherical particles, Marangoni flows, patterned substrates, and so on. However, circular patterns are usually formed because the contact line keeps circular during the droplet evaporation. Here, we show that the stretching of soft substrates strongly controls the dynamics of droplet evaporation and particle deposition through controlling the contact line motion. The findings broaden our understanding of droplet wetting and evaporation on soft and anisotropic substrates, and open the way to reshaping the coffee-ring to allow anisotropic, non-circular patterning.

**Introduction**

The evaporation of liquid droplets on solid substrates is ubiquitous in our daily experience. It has drawn significant attention from researchers and engineers, not only due to its particular importance in many fundamental physical processes, like capillarity and wetting (1, 2), phase transition (3, 4), heat and mass transfer (4-6), or self-assembly (7-9), but also due to its numerous applications in industrial processes (10, 11) and biomedical sensing (12, 13). The interaction between the droplet and the substrate is crucial for the dynamics of the droplet evaporation. On rigid substrates, contact line pinning due to the chemical or physical heterogeneities on the substrate affects the droplet evaporation dynamics by changing the local contact angle of the droplet. A strong contact line pinning force leads to a small local contact angle and thus a high local evaporative flux, which further controls the flow inside the droplet (14-16). On soft substrates, the sessile droplet deforms the substrate, creating a wetting ridge around the contact line (17-23). The wetting ridge causes viscoelastic braking of the contact line movement and thus affects the interaction between the droplet and the soft substrate (18, 24-28). It slows down droplet spreading (18), or enhance droplet evaporation (29) and condensation (30). The contact line motion on soft substrates is governed by the coupling of droplet wetting dynamics and wetting ridge dynamics (18, 22, 26, 27, 31-34).

Recently, in addition to a non-circular contact line of a droplet sitting on stretched soft substrates (35-37), anisotropic contact line motions of a droplet moving on stretched soft substrates were revealed (33, 38). This was found to result from the height gradients of the annular wetting ridge on the stretched soft substrates: the wetting ridge periodically increased and decreased in height along the contact line (33). This anisotropy in the height of the wetting ridge induces direction-dependent wetting dynamics and viscoelastic braking on stretched soft substrates (33, 38). We built on these findings, when we now study droplet evaporation on stretched soft substrates. To the best of our knowledge, no attempt has been made in this direction so far.

After the complete evaporation of a liquid droplet containing non-soluble particles on solid substrates, a circular ring deposit, i.e., the so-called coffee-ring, is formed. This phenomenon is referred to as the coffee-ring effect (14, 39, 40). Great efforts have been made to promote, suppress or even reverse this effect by manipulating the properties of the substrate, of the liquid, or of the particles and as well by controlling the environmental conditions (41-48). These attempts have stimulated the construction of different patterns via drying colloidal suspension droplets on different solid substrates (9, 11, 39, 40, 49-53). However, the shapes of the formed patterns usually only range from rings to disks, from hills to balls, no matter on rigid substrates (52-55) or on soft substrates (28, 29, 56, 57). Unfortunately, less attention was paid to the control of the coffee-ring effect in order to create non-circular patterns.

Here, we quantify how droplets dry on stretched soft substrates, i.e., how the anisotropy of the substrate induced through the stretching interacts with the droplet evaporation and particle deposition processes. The contact line starts to recede at different times and with different velocities and accelerations in the parallel and perpendicular directions to the stretching direction. This anisotropic contact line motions enable us to prepare anisotropic deposition patterns beyond the known coffee-ring by drying colloidal suspension droplets. The anisotropy of the deposition patterns can be controlled by the substrate stretching ratio. Our findings provide fundamental insight into our understanding of the contact line motions on isotropic and anisotropic soft substrates. The approach opens a simple way to prepare anisotropic non-circular patterns, e.g., for force sensor design and directional transport.

## Results

**Evaporation of pure water droplets.** Crosslinked polydimethylsiloxane (PDMS) substrates (~2 mm thick) with shear modulus $G = 16 \pm 1$ kPa were stretched by 250%. The effective stretching ratio $\lambda$ was defined as the ratio between the final and initial distances between two parallel lines perpendicular to the stretching direction on the substrates (cf. **Fig. 1 A**). Unstretched substrates with $\lambda = 1.0$ were used for comparison. A water droplet of ~0.3 μL was placed on the center of the substrate. We captured the entire droplet evaporation process either from the top-view using optical microscopy or from the side-view using two cameras placed orthogonally, as shown in **Fig. 1 A**. For the side-view imaging, camera 1 recorded the droplet shape evolution along the direction parallel to the stretching direction (i.e., the $x$ direction), and camera 2 recorded the droplet shape evolution along the direction perpendicular to the stretching direction (i.e., the $y$ direction). The evaporation experiments were performed at ambient conditions (with temperature $25.3 \pm 0.1$ °C and relative humidity $42.9 \pm 0.6$ %). Under these conditions, the droplets evaporated completely within 15 min after deposition.

Top-view images showed that the shape of the evaporating droplet changed significantly once the substrate was stretched. Whereas the droplet kept a spherical cap shape throughout its evaporation process on the unstretched substrate (**Fig. 1 B** and **Movie S1**), the droplet on the stretched substrate was elongated and showed a direction switching of its elongation direction (**Fig. 1 C** and **Movie S2**). On the stretched substrate, due to uniaxial stretching of the substrate and the resulting anisotropic surface stresses (33, 35-37), initially, the droplet was elongated by 13% along the $x$ direction of high external tension compared to along the $y$ direction. As the evaporation proceeded, the droplet took nearly a spherical cap shape at $t = 400\ s$, **Fig. 1 C**. Subsequently, the elongation direction aligned to the $y$ direction. The droplet had a spindle-like shape with its apexes (i.e., points Y and Y' in **Fig. 1 D**) becoming more and more tapered, **Fig. 1 C**. Despite these changes in the droplet shape, the total evaporation time was almost independent of the substrate stretching (*SI Appendix*, **Fig. S1**).

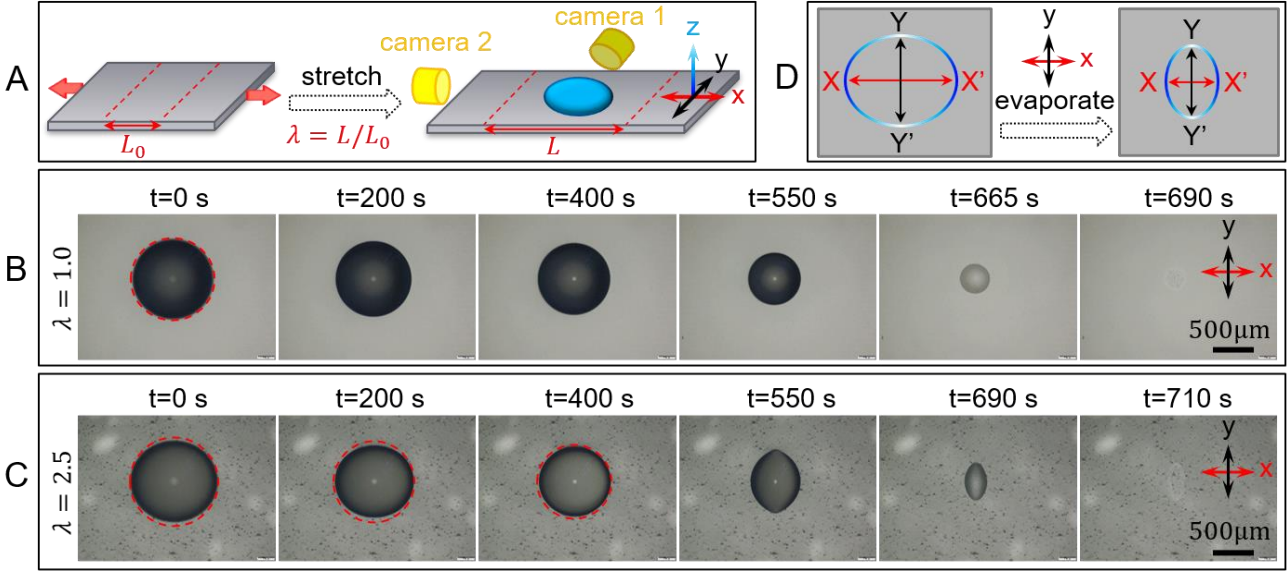

**Fig. 1.** (*A*) Schematic illustration of the effective stretching ratio, $\lambda = L/L_0$, applied to the substrate by a uniaxial stretching, and of the views of the cameras, with a coordinate indicating the directions. (*B* and *C*) Top-view snapshots of a water droplet evaporating on unstretched (*B*, $\lambda = 1.0$) and stretched (*C*, $\lambda = 2.5$) PDMS substrates with $G \approx 16$ kPa. The first column shows the initial droplet shape just after droplet deposition ($t = 0$ s) and the fifth column just before droplet collapse. The red dashed circles are for comparison to the projections or footprints of the evaporating droplets. The arrows indicate the directions parallel (the $x$ direction) and perpendicular (the $y$ direction) to the stretching direction. (*D*) Schematic illustration of the shape changes of the droplet contact line on stretched soft substrates, indicating the apexes X and X' in the $x$ direction and Y and Y' in the $y$ direction. The gradient color indicates the height gradient of the wetting ridge (33): the highest at the apexes Y and Y' and the lowest at the apexes X and X'.

**Wetting and evaporation dynamics.** To quantify the wetting and evaporation dynamics of the droplet, we measured its contact angle ($\theta_x$, $\theta_y$), the contact width ($D_x$, $D_y$, *SI Appendix*, **Fig. S2**) and the contact line velocity ($v_x$, $v_y$). The subscripts refer to the $x$ and $y$ directions, respectively. These parameters were plotted versus $t$ and $t/t_{\text{evap}}$ in **Fig. 2**. $t = 0$ s is the time after the droplet deposition and $t_{\text{evap}}$ is the total evaporation time. On the unstretched substrate, as expected, we observed $\theta_x = \theta_y$, $D_x = D_y$, and $v_x = v_y$ throughout the lifetime of the droplet, **Fig. 2 A-C**. By contrast, on the stretched substrate, these parameters are direction-dependent and show a crossover between the $x$ and $y$ directions, **Fig. 2 D-F**, due to the non-spherical cap shape of the droplet and the switching of its elongation direction (**Fig. 1 C**). In both cases, three distinct regimes, the spreading regime, the braking regime and the receding regime, can be identified successively from the evolution of the contact width (**Fig.**

**2 B** and **E**). However, the crossover time between these regimes depends on the stretching of the substrate and, in the stretched case, additionally on the orientation relative to the stretching direction. Directly after droplet deposition, in the spreading regime, the contact width increased, while the contact angle decreased with time. The contact width reached a maximum value where the corresponding contact angle can be assumed to be somehow close to the equilibrium contact angle $\theta^*$. Subsequently, the contact width decreased slowly in the following braking regime, while the contact angle kept decreasing fast with time. On the stretched substrates, the receding regime in the $x$ direction partially overlapped with the quasi-pinning regime in the $y$ direction. The contact line started to recede fast in the $x$ direction at $t/t_{\text{evap}} \approx 0.4$, whereas the contact line was further braked in the $y$ direction for additional 20% of the evaporation time. The crossover time when $D_x = D_y$ on the stretched substrate is indicated by the blue solid line in **Fig. 2 E**.

From the temporal evolution of the contact width, we determined the contact line velocity. In the braking regime, the contact line velocity is slightly smaller than the characteristic substrate relaxation velocity $v_\tau = \gamma/E\tau$, where $\gamma$ is the liquid surface tension, $E$ is the substrate elastic modulus, and $\tau$ is the characteristic substrate relaxation time (27). In our case, $v_\tau \approx 0.15 \ \mu m/s$ (with $\gamma = 72 \ \text{mN/m}$, $E = 49 \ \text{kPa}$, and $\tau \approx 10 \ s$), which is indicated by the horizontal blue dashed lines in **Fig. 2 C** and **F**. After entering the receding regime, the contact line velocity exceeds $v_\tau$, and a sub-regime with increasing acceleration is followed by increasing contact line velocity with a constant acceleration. The constant acceleration is $a = 0.0091 \ \mu m/s^2$ on the unstretched substrate (**Fig. 2 C**), and $a_x = 0.0076 \ \mu m/s^2$ in the $x$ direction and $a_y = 0.0179 \ \mu m/s^2$ in the $y$ direction on the stretched substrate (**Fig. 2 F**). We have $a_y \approx 2.0a \approx 2.4a_x$ and $a_x \approx 0.8a$. On the stretched substrate, the contact line receded later in the $y$ direction than in the $x$ direction (**Fig. 2 E**). This leads to a smaller contact angle in the $y$ direction in the receding regime (**Fig. 2 D**). Given that the capillary driving force is $\gamma(\cos\theta^* - \cos\theta)$, a larger capillary driving force is induced in this direction. This results in a larger contact line acceleration in the $y$ direction than in the $x$ direction, **Fig. 2 F**. However, although $a_y$ is 2.4 times $a_x$ on the stretched substrate, the final $v_y$ is only slightly larger than $v_x$, **Fig. 2 F**. Therefore, $D_y$ is still larger than $D_x$ before the droplet evaporates completely, **Fig. 2 E**, consistent with the shape of the evaporating droplet shown in **Fig. 1 C**.

A systematic analysis of the effects of the stretching ratio and direction on the contact line velocity in the receding regime is shown in **Fig. 3**. Usually, $v_x$ was larger than $v_y$ on the stretched substrates. At the same $t/t_{\text{evap}}$, $v_x$ monotonically increased with increasing $\lambda$ (**Fig. 3 A**), while $v_y$ showed a more complex dependence on $\lambda$ (**Fig. 3 B**). As shown in **Fig. 3 B**, for $t/t_{\text{evap}} \lesssim 0.7$, $v_y$ monotonically decreased with increasing $\lambda$. However, for $t/t_{\text{evap}} \gtrsim 0.7$, $v_y$ non-monotonically changed with $\lambda$. $v_y$ first decreased with increasing $\lambda$ and then, when $\lambda \gtrsim 2.0$, increased with increasing $\lambda$. The above results demonstrated that the contact line of an evaporating droplet on stretched soft substrates receded differently between along the stretching direction and along its perpendicular direction, i.e., anisotropically.

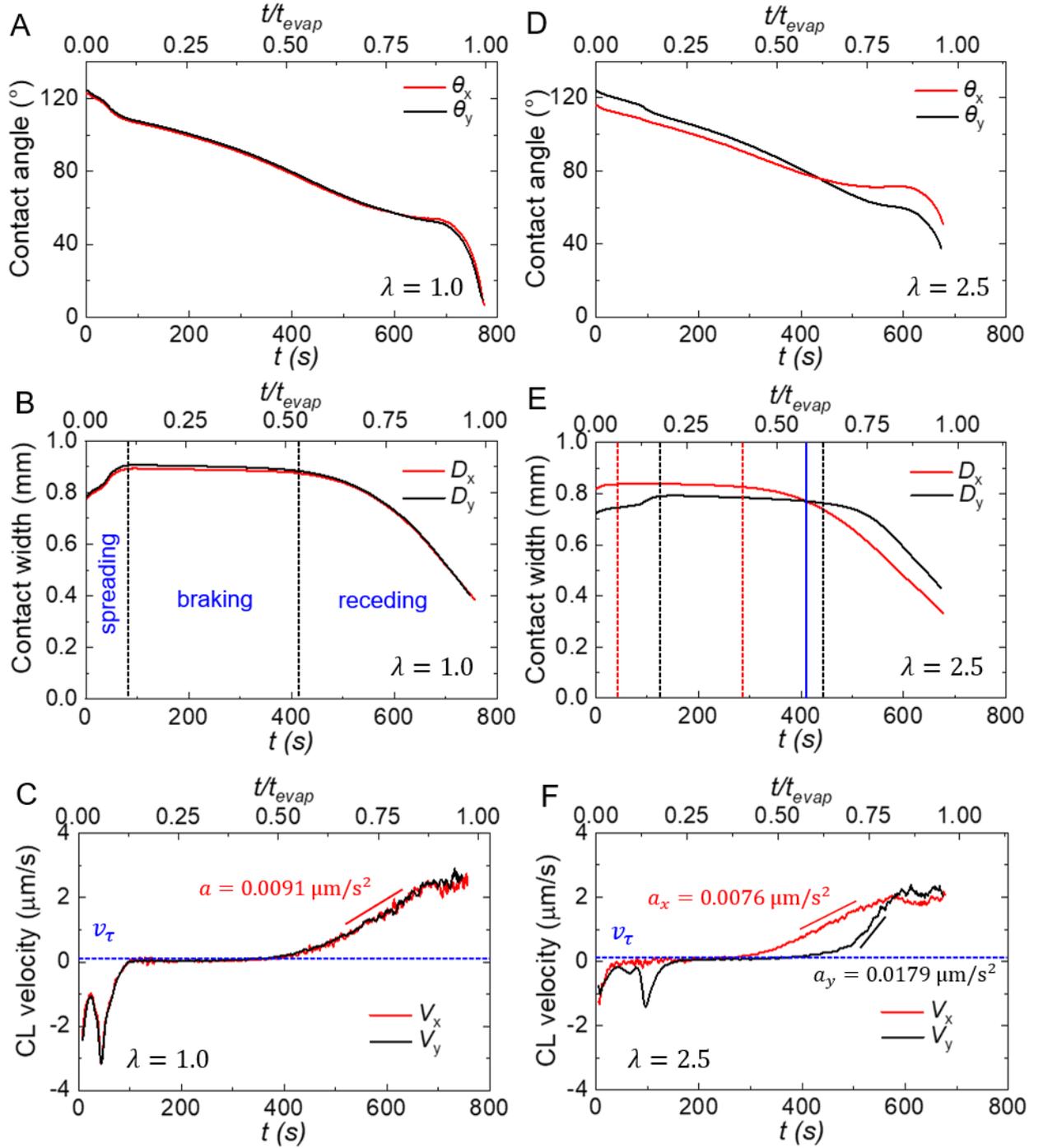

**Fig. 2.** (*A-C*) Contact angle (*A*), contact width (*B*) and contact velocity (*C*) versus $t$ and $t/t_{\text{evap}}$ for droplets evaporating on unstretched ($\lambda = 1.0$) soft substrates. (*D-F*) Contact angle (*D*), contact width (*E*) and contact velocity (*F*) versus $t$ and $t/t_{\text{evap}}$ for droplets evaporating on stretched ($\lambda = 2.5$) soft substrates. The vertical dashed lines in *B* and *E* divide the evaporation process into three regimes: spreading regime, braking regime and receding regime. The vertical blue solid line in *E* indicates the crossover time when the droplet is nearly a spherical cap shape (i.e., when $D_x \approx D_y$). The horizontal dashed lines in *C* and *F* indicate the characteristic substrate relaxation velocity $v_\tau$, and the red solid lines show the linear relations of the data in those regimes. The slopes reflect the accelerations, $a_y \approx 2.4 a_x \approx 2.0 a$.

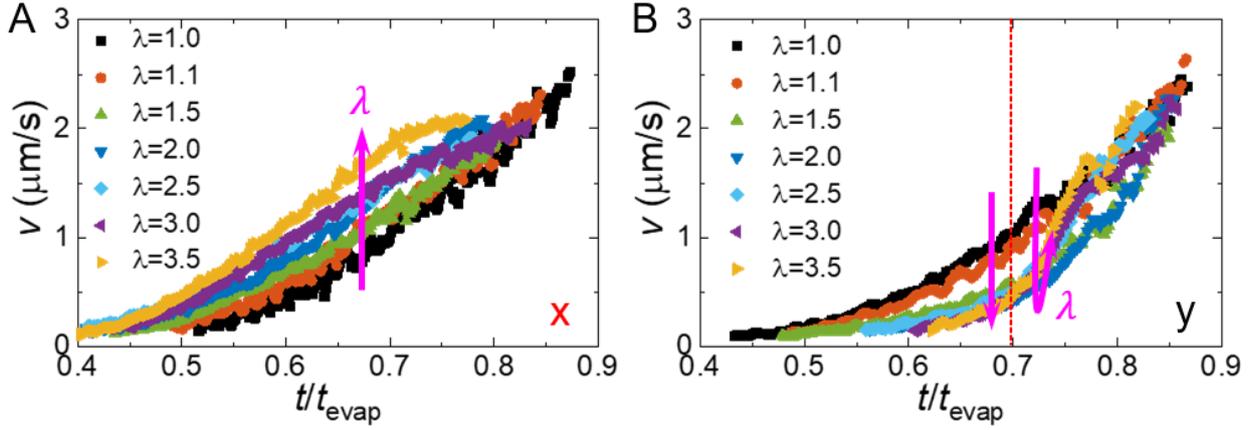

**Fig. 3.** Contact line velocity in the $x$ direction (A) and in the $y$ direction (B) versus $t/t_{\text{evap}}$ of water droplets evaporating on soft substrates stretched with different $\lambda$. The arrows indicate the increasing directions of $\lambda$.

**Theoretical models.** To theoretically understand the anisotropic contact line motions on stretched soft substrates, it is essential to analyze the energies and forces involved in this process. Thermodynamically, the released capillary free energy of the droplet is consumed partially by viscous dissipation of the liquid flow within the droplet and partially by viscoelastic dissipation due to the movement of the wetting ridge on the substrate (18, 25). Given the small contact line velocity $v \lesssim 3$ μm/s (**Fig. 3**), the viscous dissipation within the droplet is negligible and the droplet contact line motion is dominated by viscoelastic dissipation, yielding (18, 25)

$$\cos\theta^* - \cos\theta = \frac{\gamma}{2\pi G \varepsilon}\left(\frac{v}{v_0}\right)^m \qquad (1)$$

Here, $\varepsilon$ is a cutoff distance of a few nanometers near the contact line below which the behavior of the substrate is no longer linearly elastic, $v_0$ and $m$ are constants that define the damping properties of the substrate.

Alternatively, from a mechanical perspective, Karpitschka et al. developed a model to predict the onset of contact line depinning on soft substrates by balancing the capillary and elastic forces (31). They correlated the contact angle (in radians) to the contact line velocity by

$$\theta - \theta^* = \frac{2^{n-1}n}{\cos\left(\frac{n\pi}{2}\right)} \frac{\gamma \sin\theta}{\gamma_s}\left(\frac{v}{v^*}\right)^n \qquad (2)$$

where $v^*$ is the characteristic velocity scale and $n$ is an exponent depending on the stoichiometric ratio between reticulant and prepolymer of the substrate.

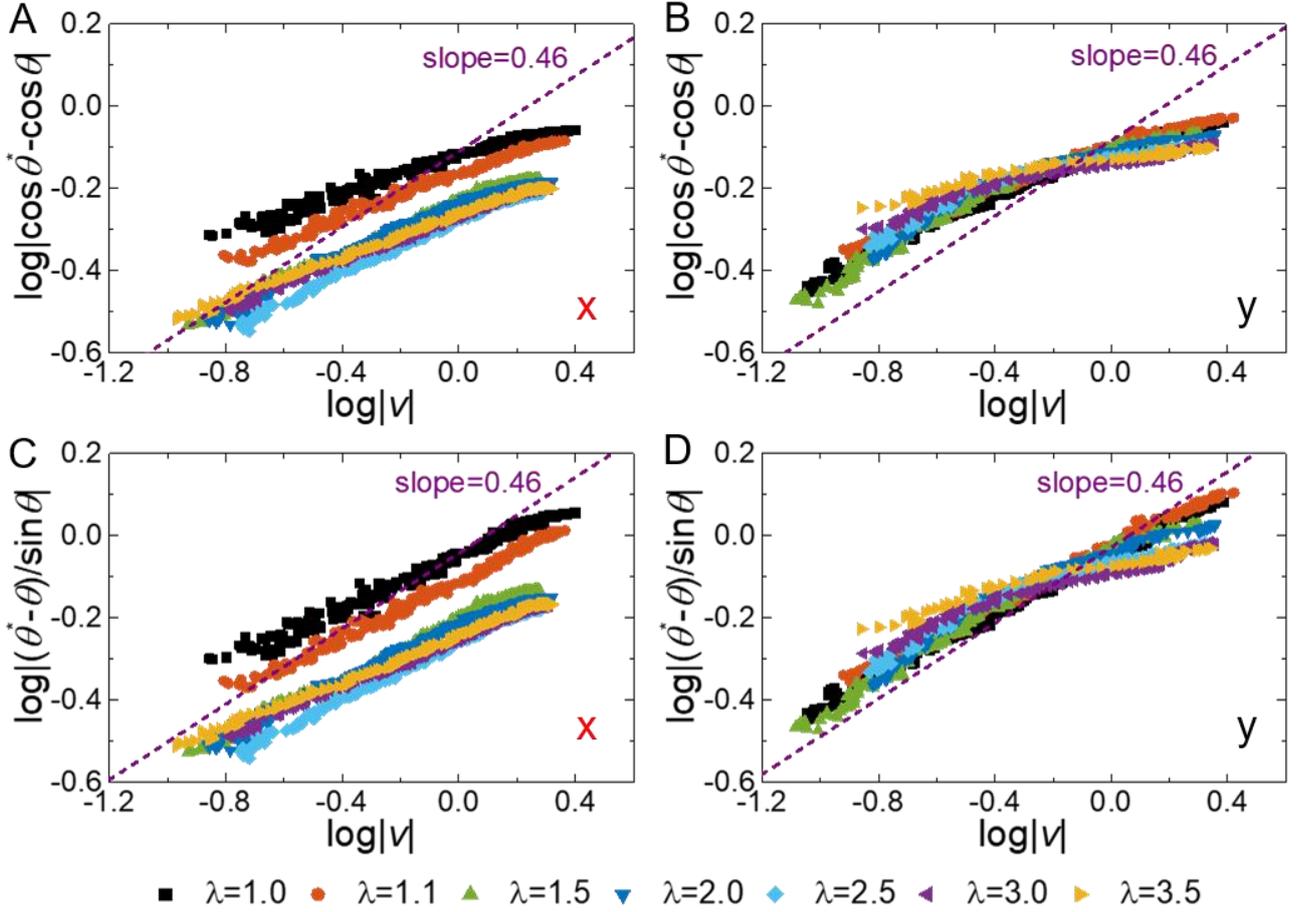

**Fig. 4.** (*A* and *B*) Variation of $\log|\cos\theta^* - \cos\theta|$ versus $\log|v|$ in the receding regimes in the $x$ direction (*A*) and in the $y$ direction (*B*). (*C* and *D*) Variation of $\log|(\theta - \theta^*)/\sin\theta|$ versus $\log|v|$ in the receding regimes in the $x$ direction (*C*) and in the $y$ direction (*D*). The dashed purple lines with a slope of $0.46$ (determined independently from rheology measurement) are guides for the eyes.

**Equations (1)** and **(2)** can be logarithmized as $\log|\cos\theta^* - \cos\theta| = m\log|v| + C_1$ and $\log\left|\frac{\theta - \theta^*}{\sin\theta}\right| = n\log|v| + C_2$, respectively, where $C_1$ and $C_2$ are constants. The logarithmic forms were used to analyze the receding regimes of the experimental results, **Fig. 4**. Both models are compatible with our data. For both models, a nearly linear relation can be found in the $x$ direction, but not in the $y$ direction. Therefore, we only determined the exponents $m$ and $n$ from the data in the $x$ direction at different $\lambda$ (**Fig. 4 *A*** and ***C***). Both exponents are approximately 0.3, which are smaller than that determined from the rheology measurement (*SI Appendix*, **Fig. S3**), 0.46, and they show no dependence on $\lambda$, *SI Appendix*, **Fig. S4**.

**Non-circular deposition patterns.** The above results show that stretching a soft substrate can induce anisotropic contact line motions. This offers the possibility to create patterns beyond the circular coffee-ring. As shown in the last column in **Fig. 1 *B*** and ***C***, the footprints of the deposits of impurities in water replicated the contact line geometries of the droplets just before they completely dried (the fifth column in **Fig. 1 B** and **C**). The topographies of the deposition patterns are more clearly shown in the confocal microscopy images (*SI Appendix*, **Fig. S5,** the first column). On the unstretched substrate, a circular deposition pattern, i.e., the so-called coffee-ring, was formed (the last column in **Fig. S5 A**). By contrast, a non-circular deposition pattern was formed on the stretched substrate (the last column in **Fig. S5 B**). The elongation direction of the non-circular deposition pattern aligned along the $y$ direction, which is orthogonal to the stretching direction.

For a detailed investigation of the deposition patterns, water droplets with dispersed silica nanoparticles with a diameter of $r = 237 \pm 14$ nm were dried on the soft substrates. The deposition patterns were characterized by confocal microscopy imaging. Firstly, three different concentrations of the nanoparticles, $c = 0.005$ wt%, $0.02$ wt% and $0.1$ wt%, were examined. The overall shapes of the deposition patterns were not affected by the addition of the nanoparticles or by their concentrations, only the rim thickness (in the horizontal plane) and the rim height (orthogonal to the horizontal plane) increased with increasing $c$ (*SI Appendix*, **Fig. S5**). Whereas the circular rings formed on the unstretched substrates are homogeneous in height ($H_0$) at each concentration, the non-circular deposition patterns formed on the stretched substrates exhibit height gradients along the rim. The rim is the highest at the apexes Y and Y' and the lowest at the apexes X and X', which is schematically illustrated in the last panel in **Fig. 5 A**.

In a further study, we chose the particle concentration $c = 0.02$ wt% to systematically investigate how substrate stretching affects the morphology of the pattern. The stretching ratio applied to the substrate was tuned in the range of $\lambda = 1.0 - 4.0$. **Figure 5 A** shows that the circular coffee-ring formed on the unstretched substrates became to be elongated along the $y$ direction on the stretched substrates. The finer morphology of the elongated non-circular deposition patterns depends on $\lambda$. For a quantitative analysis, we measured the widths (i.e., in the $x$ direction, $W_x$, and in the $y$ direction, $W_y$) and heights (i.e., the average height at the apexes X and X', $H_x$, and the average height at the apexes Y and Y', $H_y$) of the patterns from their cross-sectional profiles (*SI Appendix*, **Fig. S6**). As shown in **Fig. 5 B** and **C**, the widths and heights in both directions non-monotonically changed with $\lambda$. $W_y$, $H_x$ and $H_y$ initially increased and then decreased with increasing $\lambda$, while $W_x$ showed the opposite tendency. $W_x$ had the minimum and $H_x$ and $H_y$ had the maximum values at $\lambda \approx 2.5$, while $W_y$ had the maximum value at $\lambda \approx 1.5$. However, the aspect ratios $W_y/W_x$ and $H_y/H_x$ both increased with increasing $\lambda$ when $\lambda \lesssim 2.0$. With further increasing $\lambda$, $W_y/W_x$ decreased, while $H_y/H_x$ saturated (**Fig. 5 D**). Unfortunately, a circular ring cannot recur before the break of the substrate. Note that the non-monotonic change of the pattern geometry with $\lambda$ cannot be ascribed to the yielding point of the substrate because no unambiguous yielding point can be identified for the substrates used in our experiments (*SI Appendix*, **Fig. S7**). The critical $\lambda$ value at which the aspect ratios of the patterns showed different tendencies in **Fig. 5 D** matches the critical $\lambda$ value at which $v_y$ non-monotonically changed with $\lambda$ in **Fig. 3 B**.

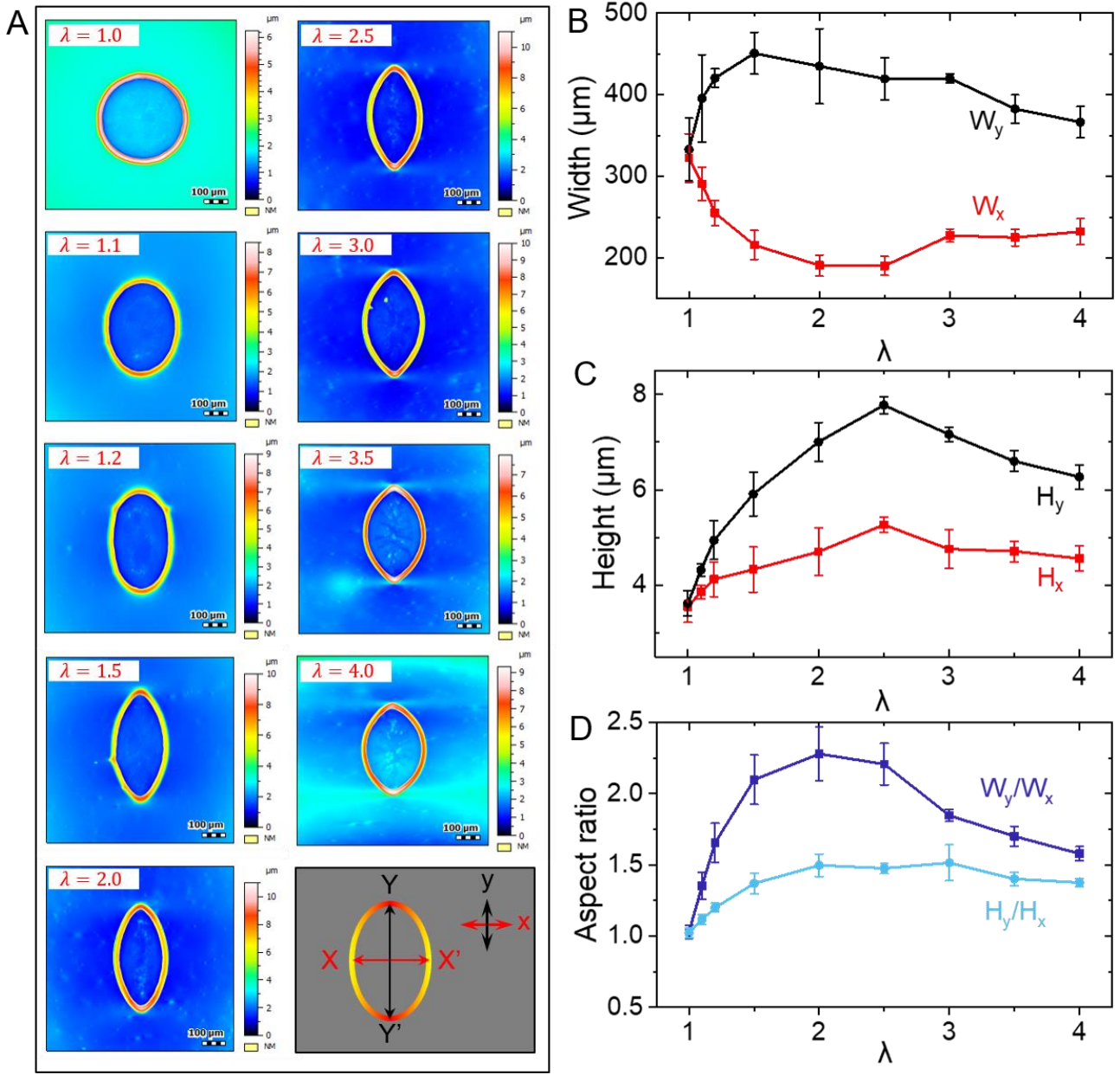

**Fig. 5.** (*A*) Morphologies of the deposition patterns formed at different $\lambda$ with $c = 0.02$ wt%. The last panel schematically illustrates the height gradients at the rim of the non-circular deposition pattern. The rim is the highest at the apexes Y and Y' and the lowest at the apexes X and X'. (*B*) Widths and (*C*) heights of the deposition patterns versus $\lambda$. (*D*) The aspect ratios $W_y/W_x$ and $H_y/H_x$ versus $\lambda$.

In contrast to the distinct change between the deposition patterns formed on unstretched and stretched soft PDMS substrates ($G \approx 16$ kPa), we demonstrated that the substrate stretching does not induce a significant change of the final pattern on a rigid PDMS substrate of $G \approx 667$ kPa. On such rigid PDMS substrate stretched at $\lambda = 1.5$, the aspect ratio of the final pattern is $W_y/W_x \approx 1.05$ and without distinguishable height gradient, even without distinguishable difference from the unstretched case (*SI Appendix,* **Fig. S8**). However, on the soft PDMS substrate at the same stretching ratio, we have $W_y/W_x \approx 2.09$ and $H_y/H_x \approx 1.36$ (**Fig. 5 *D***). These results demonstrate that the substrate stiffness plays a critical role on the geometry of the patterns formed on stretched substrates. This can be ascribed to the formation of the wetting ridge, which induces viscoelastic dissipation (18, 24-28). On the unstretched soft PDMS substrate, the height of the wetting ridge scales as $\gamma/G \approx 4.5\ \mu m$. This value decreases to $0.1\ \mu m$ on the unstretched rigid PDMS substrate, with a further decrease under stretching (35, 58, 59). Therefore, the rigid PDMS substrate behaved only with subtle viscoelastic dissipation, leading to almost isotropic contact line motion and hence almost circular patterns without distinguishable height gradient (*SI Appendix*, **Fig. S8**).

**Discussion**

The anisotropic contact line receding motions of the evaporating droplet on the stretched soft substrate should be ascribed to the uniaxial substrate deformation. The droplet sitting on the soft substrate deforms the substrate (17-24, 60), creating a wetting ridge below the contact line. On the non-stretched substrate, the wetting ridge is uniform in height (20-23). Applying an external stretching to the substrate changes not only the height of the wetting ridge (33, 35, 58, 59), but also the mechanical properties of the substrate (35, 58, 61). The wetting ridge becomes shallower along the stretching direction than along the perpendicular direction, showing a height gradient (i.e., a non-monotonic but regular height variation with two maxima at the apexes Y and Y' and two minima at the apexes X and X', as schematically illustrated in **Fig. 1 *D***) (33, 35). Xu et al. (35, 58) found that the height of the wetting ridge decreased with increasing the stretching, with a more pronounced change at the position X (same to the position X') than at the position Y (same to the position Y'). The local strain in the radial direction, which is the sum of the strain produced by the external stretching and the extra strain due to the presence of the wetting ridge (58), increased at the position X but decreases at the position Y (*SI Appendix*, **Fig. S9 *A***) as the applied stretching increased (35). We linearly fitted data from Xu et al. (35) acquired at the positions X and Y and calculated the surface stresses by $\Upsilon = \gamma_s + \Lambda(\lambda - 1)$, where $\Lambda = 126\ \text{mN/m}$ is the surface elastic constant (35, 58). $\Upsilon$ at the position X, $\Upsilon_x$, is always larger than at the position Y, $\Upsilon_y$. $\Upsilon_x$ increased and $\Upsilon_y$ decreased with increasing $\lambda$. However, $\Upsilon_y$ showed a sign change at $\lambda \approx 2.0$ (*SI Appendix*, **Fig. S9 *B***). On the one hand, a larger surface stress means a weaker viscoelastic dissipation in the wetting ridge and thereby a weaker viscoelastic braking effect, which results in a higher contact line mobility (27, 33, 38). Therefore, the contact line can more easily move in the $x$ direction than in the $y$ direction. On the other hand, at small $\lambda$ ($\lambda \lesssim 2.0$), $\Upsilon_y$ is positive and decreases as $\lambda$ is increased, suggesting that the substrate behaves softer at the position Y when increasing $\lambda$. Thus, a stronger viscoelastic braking effect causes a smaller contact line mobility. Oppositely, for $\lambda \gtrsim 2.0$, $\Upsilon_y$ is negative. The substrate behaves stiffer at the position Y where the contact line mobility would be increased when $\lambda$ increases (*SI Appendix*, **Fig. S9 *B***). Therefore, the sign change of $\Upsilon_y$ could be correlated to the monotonic to non-monotonic change of $v_y$ with $\lambda$ in the receding regime of the droplet evaporation process (**Fig. 3 *B***).

The anisotropic contact line motions resulted in non-circular deposition patterns with periodic height gradients. Nanoparticles preferentially accumulate in the regions of a large contact line curvature where the higher evaporation rate leads to a stronger fluid flow and thus drags more particles to this region (14, 15). On the non-stretched substrate, the contact line receded isotropically (**Fig. 1 *B***), leading to a uniform curvature of the contact line and a circular ring pattern with uniform height along the ring. By contrast, on the stretched substrate, the orientation-dependent contact line velocity caused a curvature of the contact line much higher at the positions Y and Y' than at the positions X and X'. Consequently, the evaporation flux near the contact line changed periodically and with gradients along the contact line. The nanoparticles were transported to the contact line near the positions Y and Y' due to the stronger evaporative flux there, while fewer particles accumulate at the contact line near the positions X and X'. Upon the collapse of the droplet at the very late evaporation stage, nanoparticles were left over at the contact line of the droplet just before it collapsed. Since the contact line was elongated, the rim of the pattern replicated this shape, forming an elongated deposition pattern with periodic height gradients. Local scanning electron microscopy (SEM) images confirmed that most of the nanoparticles were assembled densely at the rim, some nanoparticles were randomly distributed within the rim, but almost no nanoparticles were seen outside of the rim (*SI Appendix*, **Fig. S10**).

The dynamics of wetting and evaporation of a droplet on soft substrates is far more complex due to the substrate deformation than on ordinarily rigid substrates. Applying an external stretching to the soft substrates additionally induces height gradients along the wetting ridge, significantly altering the wettability, the contact line motion, the droplet evaporation and the particle deposition. In this study, we showed that droplets evaporated with anisotropic wettability and contact line motions in the directions parallel and perpendicular to the stretching direction on stretched soft substrates. The anisotropy was controlled by the stretching ratio. This led to the formation of elongated non-circular deposition patterns with height gradients along the rim. All these phenomena were correlated to the dependences of the local surface stresses (i.e., near the contact line) on the stretching direction and the stretching ratio. The knowledge gained in this study is essential for understanding the wetting and evaporation dynamics of droplets on soft and on anisotropic substrates as well as for constructing anisotropic patterns, which may have potential applications in designing force sensors and bendable and transparent electrodes.

**Methods**

**Preparation of soft substrates.** Soft substrates were made of polydimethylsiloxane gel (PDMS, Sylgard 184 silicone elastomer, Dow Corning) consisting of silicone base and crosslinking agent with a mass ratio of 40:1. Substrates with the stochiometric ratio 10:1 were used as rigid substrates. The mixtures were thoroughly mixed in a conditioning mixer (ARE-250, THINKY) at 2000 rpm for 2 min followed by 2200 rpm for 2 min. After pouring into a petri dish, the mixtures were degassed for 30 min before curing in a vacuum oven (Heraeus Instruments Vacutherm, Thermo Scientific) at 80 °C overnight. The thickness of the resultant substrates was around 2 mm. With respect to the aging effect, the substrates were always used within one week after preparation.

**Rheology measurements.** The storage ($G'$) and loss moduli ($G''$) of the crosslinked PDMS 40:1 substrate were investigated with an Anton Paar MCR301 rheometer using a plate-plate geometry of diameter 50 mm. The range of the angular frequency was $\omega = 0.1 - 100$ rad/s. $G'$ and $G''$ versus $\omega$ were fitted using the relation $G' + iG'' = G[1 + (i\tau\omega)^n]$ (31), (*SI Appendix*, **Fig. S3**). Fitting results from four measured samples gave $G = 15 \pm 2$ kPa, and $n = 0.46 \pm 0.01$.

**Tensile measurements.** The material properties under uniaxial tensile loading were determined for each material according to DIN EN ISO 527-2 on three samples with Type 5A geometry. The tests were performed on a Zwick tensile test rig with a 100 N load cell and pneumatic clamp jaws. Strain measurement was performed using a 2D optical strain measurement system. Stress-strain curves were measured until failure of the substrates (*SI Appendix*, **Fig. S7**). The elastic modulus $E$ was determined by linear regression of the data in the strain range of $0.05 - 0.3$. From the determined $E$, the shear modulus ($G$) was calculated based on the relation $G = E/2(1 + \nu)$ with the Poisson's ratio $\nu = 0.5$. For PDMS substrates with mass ratios of 10:1 and 40:1, the statistical results gave $G = 667 \pm 78$ kPa and $G = 16 \pm 1$ kPa, respectively, and the substrates broke at $\lambda = 1.94 \pm 0.02$ and $\lambda = 3.99 \pm 0.01$, respectively.

**Stretching of the substrates.** The PDMS substrates were pre-stretched before droplet deposition. Two parallel lines perpendicular to the stretching direction with an initial distance $L_0$ were marked on the substrate. After gently stretching the substrate, a final distance $L$ between the two marker lines was reached. $L$ was at least 5 times the droplet contact width. The effective substrate stretching ratio was defined as $\lambda = L/L_0$ (**Fig. 1 A**). The applied stretching ratios were in the range of $\lambda = 1.0 - 4.0$. We emphasize that the final distance did not change after the droplet evaporation and the deposition pattern characterization, excluding the effect of the possible relaxation of the substrate stretching on the droplet evaporation dynamics and the patterns geometries.

**Preparation of colloidal suspensions.** Monodisperse silica ($SiO_2$) nanoparticles with a diameter of $r = 237 \pm 14$ nm were suspended in water (purified with Smart2Pure water purification system, Thermo Scientific). An SEM image of the nanoparticles was provided in *SI Appendix*, **Fig. S11**. Detailed synthesis process was described elsewhere (62). We prepared suspensions with final silica nanoparticle concentrations of 0 wt% (i.e., without adding silica nanoparticles), 0.005 wt%, 0.02 wt% and 0.1 wt%. The suspensions were used within three days after preparation and were treated in ultrasonic bath with water for 5 min before each use.

**Droplet evaporation experiments.** A syringe with a needle of $0.24$ mm in outer diameter was mounted on a contact angle measuring device (DataPhysics OCA35L, Germany) to generate a pendant droplet (~$0.3$ μL). The substrate was moved up to touch the pendant droplet. When moving back the substrate the droplet detached from the needle and rested on the substrate. The droplet dried within $15$ min. The entire droplet evaporation process was recorded after droplet deposition as soon as possible, either from top-view using optical microscopy or from side-view using two cameras placed orthogonally (**Fig. 1 *A***). For top-view imaging, the sample was quickly transferred to the optical microscope (Olympus BX51), then the recording was started after droplet deposition as soon as possible. The optical microscope was adjusted and focused before each measurement. For the side-view imaging, camera 1 (a Nikon digital camera, D7500) recorded the droplet shape evolution in the direction (i.e., the $x$ direction) parallel to the stretching direction, while camera 2 (built in the contact angle measuring device) recorded the droplet shape evolution in the perpendicular direction (i.e., the $y$ direction). The evaporation experiments were performed in ambient conditions, with temperature $T = 25.3 \pm 0.1$ °C and relative humidity $RH = 42.9 \pm 0.6$ %.

**Pattern characterization.** After the droplet dried, the residual structures on the substrates were characterized by a 3D scandisk confocal microscope (NanoFocus, μSurf explorer, Oberhausen, Germany) and SEM (Zeiss NOEN40 SEM/FIB system). The obtained images allowed characterizing the morphology of the deposition patterns formed.

## Acknowledgements

We greatly acknowledge Vincent Körber for 3D printing of the stretching stage, Stefan Michel for performing the rheology measurements, Holger Scheibner for performing the tensile measurements, and Astrid Drechsler for providing technical support in Nanofocus imaging. This study was funded by the Deutsche Forschungsgemeinschaft (DFG, German Research Foundation): Project ID 265191195 - SFB 1194 and 456180046.

## Author contributions

B.Z. conceived research and designed experiments; Y.D. and B.Z. performed experiments and analyzed data; B.Z. and G.K.A. interpreted data and worked on the theoretical models; K.U. performed the tensile measurements; B.Z., G.K.A., E.B., Y.D., L.C. and J.G. discussed results; B.Z. wrote the manuscript with comments and input from G.K.A., E.B., and others.

## Additional information

**Competing interests**: The authors declare no competing interest.

**Supporting Information**: This article contains supporting information.

**Data Availability**. Raw data have been deposited in Zenodo (https://doi.org/10.5281/zenodo.7385246) (63).

**Supporting information**

**How droplets dry on stretched soft substrates**

Yixuan Du[a], Elmar Bonaccurso[b], Jianwei Guo[c], Kai Uhlig[a], Longquan Chen[d],

Binyu Zhao[a,*] & Günter K. Auernhammer[a]

[a] Leibniz Institute of Polymer Research Dresden, Dresden 01069, Germany
[b] Airbus Central R&T, Materials, Munich 81663, Germany
[c] School of Mechanics and Aerospace Engineering, Southwest Jiaotong University, Chengdu 610031, China
[d] School of Physics, University of Electronic Science and Technology of China, Chengdu 610054, China

\* To whom correspondence should be addressed. E-mail: binyuzhao2014@sina.com.

1. **Supplementary Videos**

**Movie S1.** A top-view video of a water droplet drying on the unstretched substrate.

**Movie S2.** A top-view video of a water droplet drying on the 250% stretched substrate.

2. **Supplementary Figures**

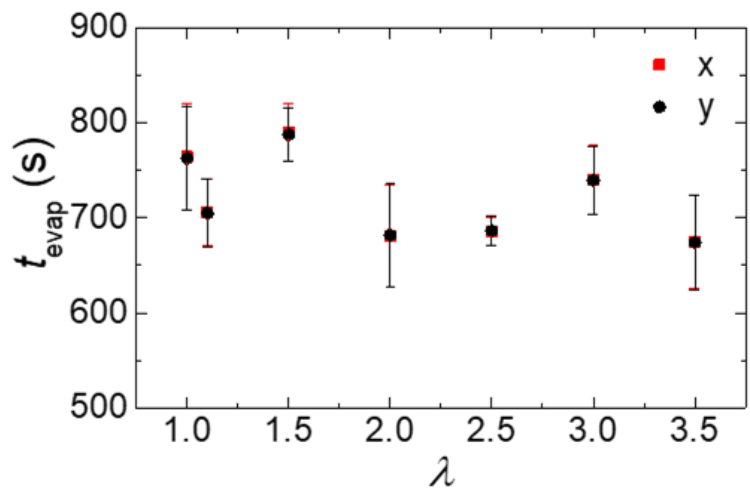

**Fig. S1.** Total evaporation time (in the $x$ and the $y$ directions) of water droplets evaporating on soft substrates at different stretching ratios.

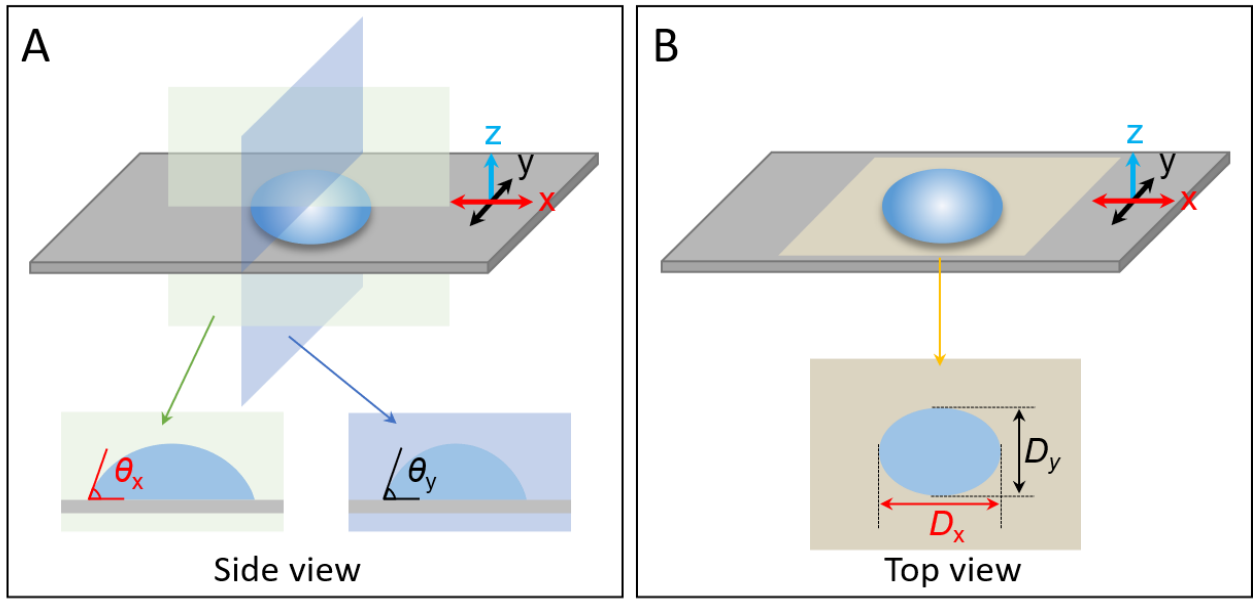

**Fig. S2.** (*A*) Side view sketch indicates the contact angles in the $x$ ($\theta_x$) and $y$ ($\theta_y$) directions of a droplet on stretched soft substrates. (*B*) Top view sketch indicates the width of the projected contact line in the $x$ ($D_x$) and $y$ ($D_y$) directions of a droplet on stretched soft substrates.

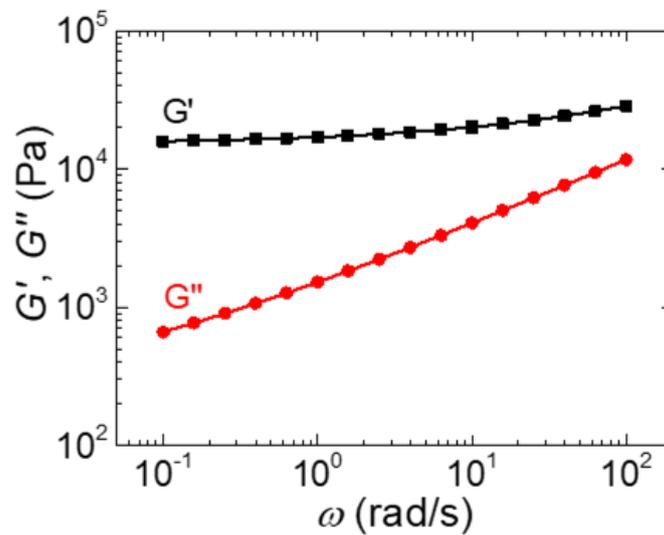

**Fig. S3.** A representative rheology measurement of PDMS 40:1 substrate. The storage ($G'$) and loss moduli ($G''$) versus the angular frequency ($\omega$) were fitted (solid lines) using the relation $G' + iG'' = G[1 + (i\tau\omega)^n]$, giving $G = 15 \pm 2$ kPa, and $n = 0.46 \pm 0.01$.

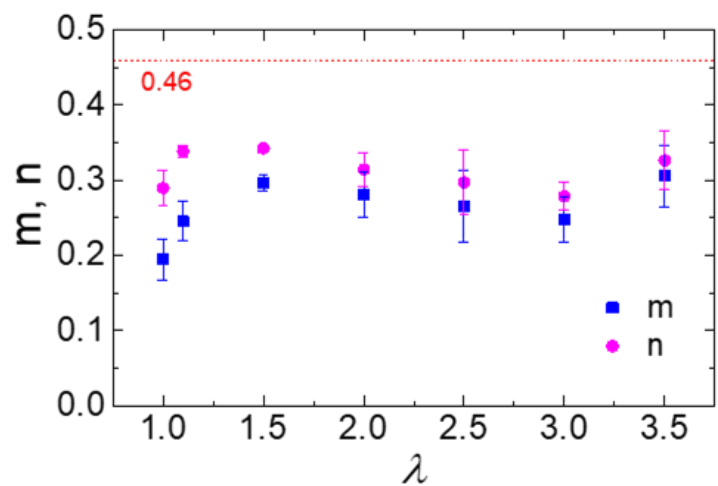

**Fig. S4.** The exponents $m$ and $n$ versus $\lambda$ from the linear fitting of the data in the $x$ direction in receding regime. The horizontal dashed line indicates the exponent determined independently from rheology measurement, $0.46$.

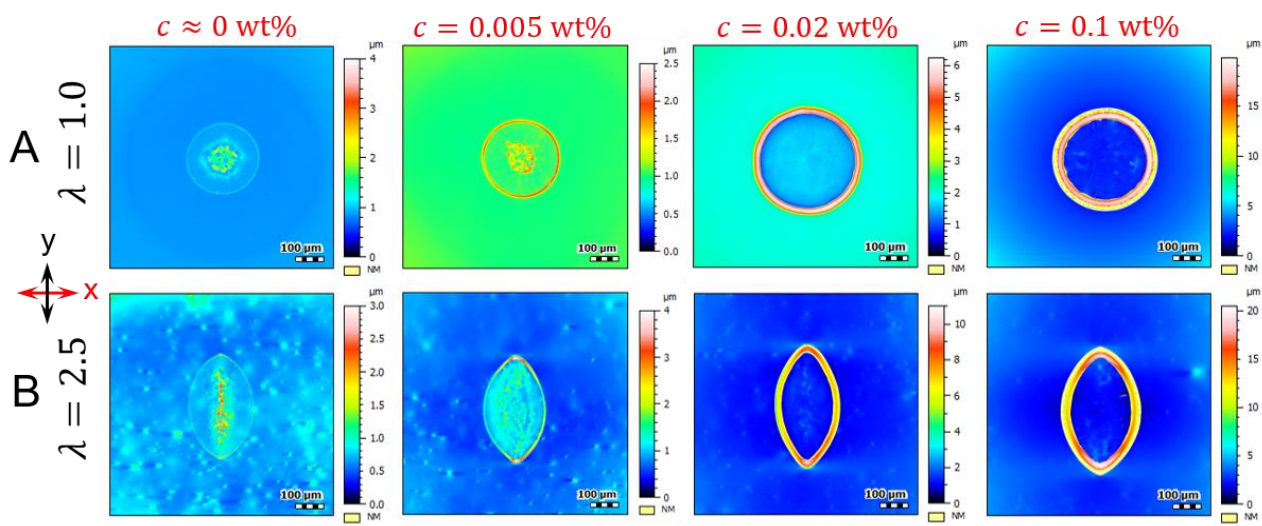

**Fig. S5.** Effects of nanoparticle concentration on the pattern geometry on soft substrates at (A) $\lambda = 1.0$ and (B) $\lambda = 2.5$. The nanoparticle concentrations are indicated at the top of each column.

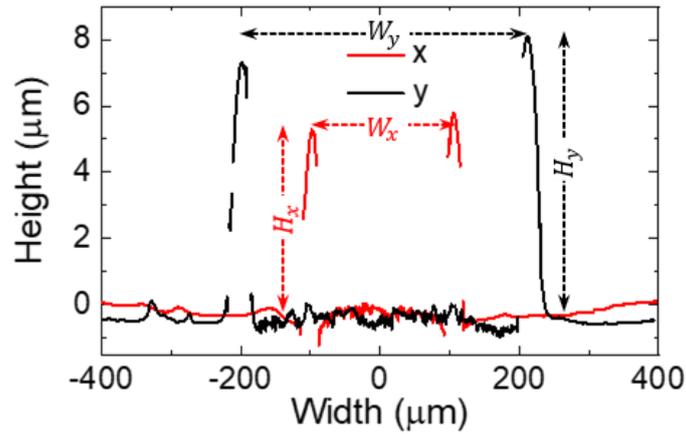

**Fig. S6.** Cross-sectional profile of a representative elongated non-circular deposition pattern ($\lambda = 2.5$). The widths ($W_x$, $W_y$) and heights ($H_x$, $H_y$) are indicated. $H_x$ and $H_y$ are given by the average heights measured at both sides of the rim in the $x$ and $y$ directions, respectively.

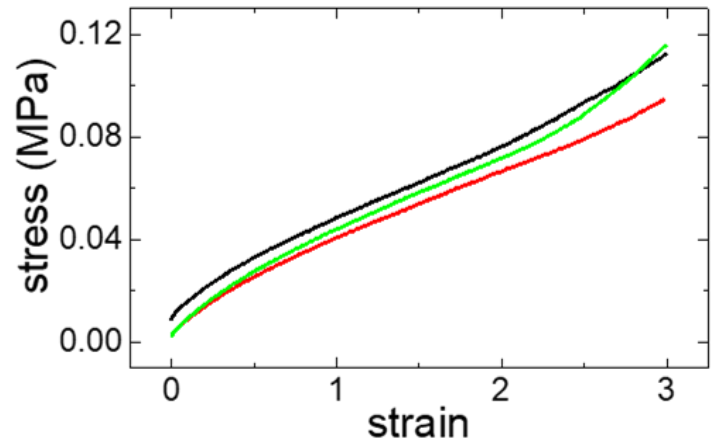

**Fig. S7.** Stress-strain curves from the uniaxial tensile strength measurements of PDMS 40:1 substrate. Different colors represent different samples tested.

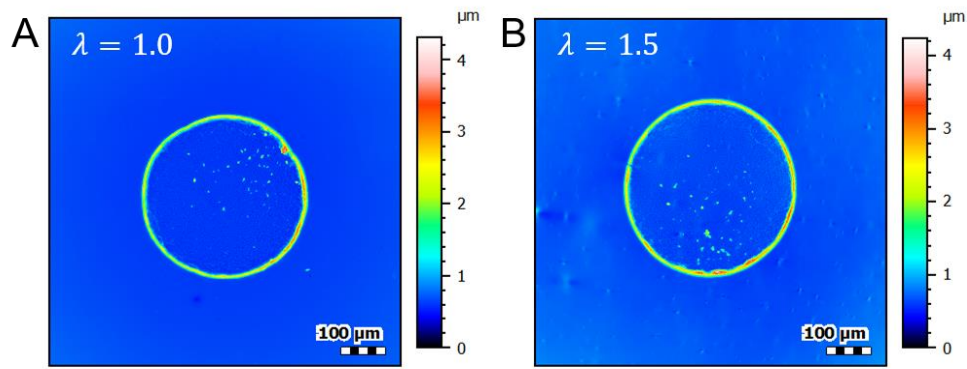

**Fig. S8.** Topographies of the patterns formed on PDMS 10:1 substrate at (A) $\lambda = 1.0$ and (B) $\lambda = 1.5$. The concentration of the nanoparticles suspended in water is $c = 0.02$ wt%.

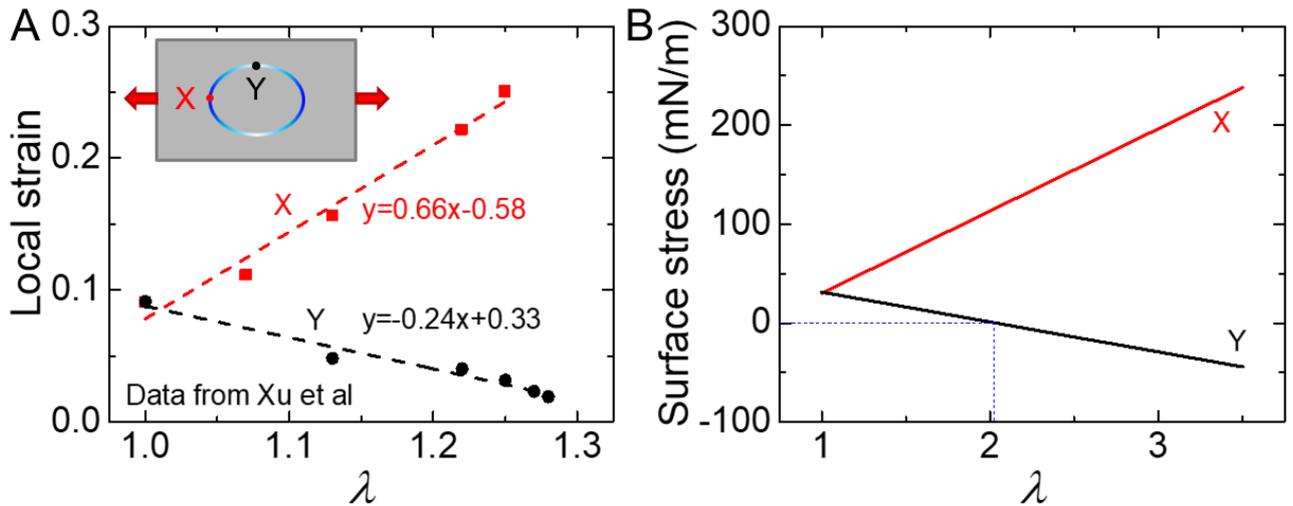

**Fig. S9.** (*A*) The local strain versus $\lambda$ at the positions X and Y (see the inset sketch) at the wetting ridge. The experimental data points were extracted from Ref. (35). The applied strain had been represented by $\lambda$. The dashed lines show the linear fits. (*B*) Calculated surface stress versus $\lambda$ at the positions X and Y at the wetting ridge. A sign change of the surface stress at the position Y versus $\lambda$ was found at $\lambda \approx 2.0$.

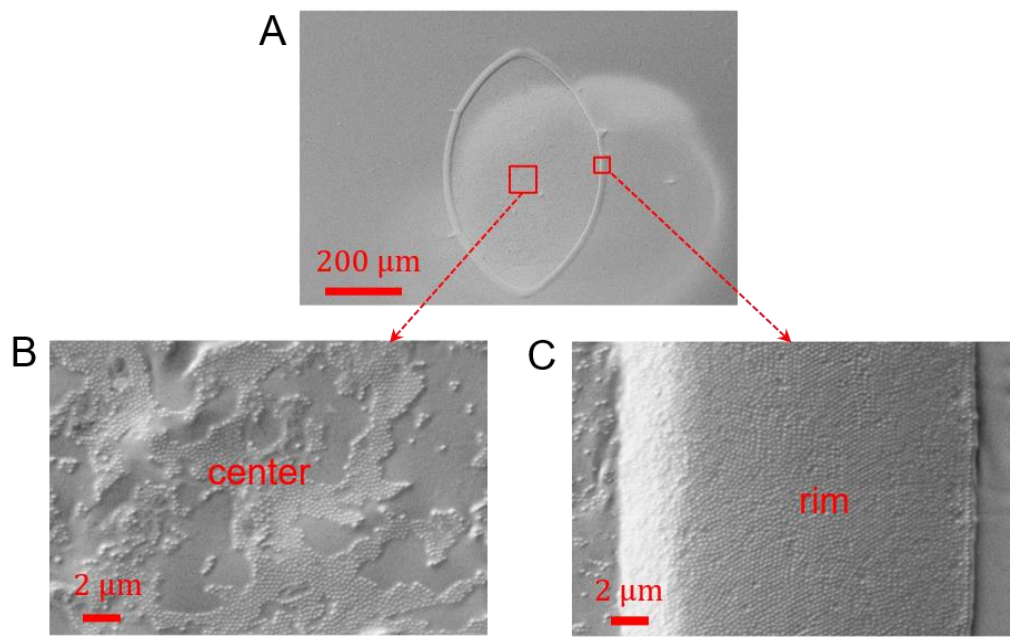

**Fig. S10.** SEM images of an elongated non-circular deposition pattern formed on a stretched PDMS 40:1 substrate. (*A*) The overall pattern. (*B*) In the center of the pattern. (*C*) On the rim.

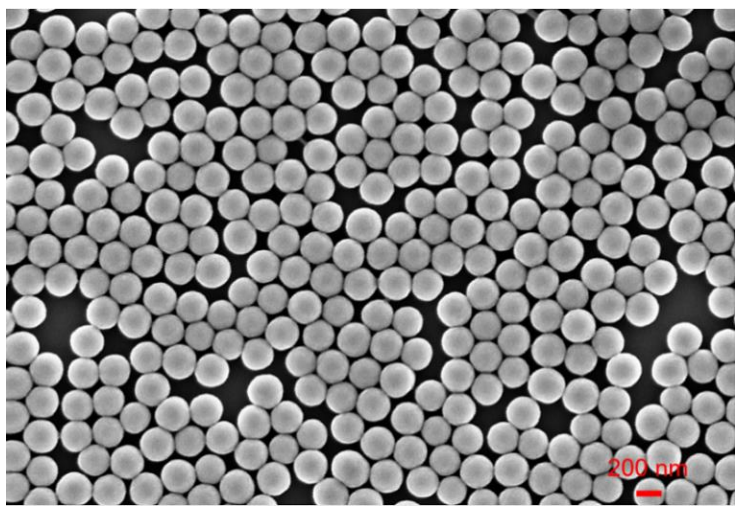
**Fig. S11.** An SEM image of the silica nanoparticles used in this study. The statistical diameter of the nanoparticles is $237 \pm 14$ nm.